\documentclass[letterpaper]{article} % DO NOT CHANGE THIS
\usepackage[draft]{aaai2026}  % DO NOT CHANGE THIS
\usepackage{times}  % DO NOT CHANGE THIS
\usepackage{helvet}  % DO NOT CHANGE THIS
\usepackage{courier}  % DO NOT CHANGE THIS
\usepackage[hyphens]{url}  % DO NOT CHANGE THIS
\usepackage{graphicx} % DO NOT CHANGE THIS
\usepackage{natbib}  % DO NOT CHANGE THIS AND DO NOT ADD ANY OPTIONS TO IT
\usepackage{caption} % DO NOT CHANGE THIS AND DO NOT ADD ANY OPTIONS TO IT
\usepackage{algorithm}
\usepackage{algorithmic}

\usepackage[utf8]{inputenc}
\usepackage{tikz}
\usepackage{booktabs}
\usepackage{multirow}
\usepackage{pdfpages} 
\usepackage{graphicx}
\usepackage{cuted}

\usepackage{newfloat}
\usepackage{listings}
\DeclareCaptionStyle{ruled}{labelfont=normalfont,labelsep=colon,strut=off} % DO NOT CHANGE THIS
\floatstyle{ruled}
\newfloat{listing}{tb}{lst}{}
\floatname{listing}{Listing}
\title{From Verification Burden to Trusted Collaboration: Design Goals for LLM-Assisted Literature Reviews}
\author{
    Brenda Nogueira\textsuperscript{\rm 1}, %\thanks{With help from the AAAI Publications Committee.}\\
    Werner Geyer\textsuperscript{\rm 2},
    Andrew Anderson\textsuperscript{\rm 2},
    Toby Jia-Jun Li\textsuperscript{\rm 1},
    Dongwhi Kim\textsuperscript{\rm 1}, %\\
    Nuno Moniz\textsuperscript{\rm 1}, %\equalcontrib,
    Nitesh V. Chawla\textsuperscript{\rm 1}%\equalcontrib
}
\affiliations{
    \textsuperscript{\rm 1} University of Notre Dame\
    \textsuperscript{\rm 2} IBM Research\\

}

\usepackage{bibentry}
\begin{document}

\maketitle

\begin{abstract}
Large Language Models (LLMs) are increasingly embedded in academic writing practices. Although numerous studies have explored how researchers employ these tools for scientific writing, their concrete implementation, limitations, and design challenges within the literature review process remain underexplored. In this paper, we report a user study with researchers across multiple disciplines to characterize current practices, benefits, and \textit{pain points} in using LLMs to investigate related work. We identified three recurring gaps: (i) lack of trust in outputs, (ii) persistent verification burden, and (iii) requiring multiple tools. This motivates our proposal of six design goals and a high-level framework that operationalizes them through improved related papers visualization, verification at every step, and human-feedback alignment with generation-guided explanations. Overall, by grounding our work in the practical, day-to-day needs of researchers, we designed a framework that addresses these limitations and models real-world LLM-assisted writing, advancing trust through verifiable actions and fostering practical collaboration between researchers and AI systems. 
\end{abstract}


\section{Introduction}

The integration of Large Language Models (LLMs) into academic practices has seen a rapid and substantial increase, particularly since the release of ChatGPT in late 2022~\citep{liang2024mapping,Kannan_2024,Quthor_2024}. This accelerated adoption is evident across various scientific disciplines, with Computer Science papers exhibiting the most significant and fastest growth in LLM usage, reaching up to 17.5\% of abstracts and 15.3\% of introductions by February 2024~\citep{liang2024mapping}. These tools also used in scientific processes such as solving mathematical problems, assisting with proofs, generating code, and discovering patterns in large datasets~\citep{liang2024mapping,Kannan_2024,si2024can}. The widespread diffusion of these technologies is transforming both how research is conducted and disseminated~\citep{Kannan_2024}. %Despite this widespread integration and growing role of AI across scientific practices, its concrete impact on writing and reporting progress remains underexplored—particularly regarding how researchers utilize LLMs, the advantages they find, and the challenges they encounter.

%Despite this widespread integration and the potential of AI for scientific writing, significant gaps remain in understanding how researchers interact with LLMs in this context, what the benefits are, and the challenges when crafting scientific reports.

%While LLMs offer notable advantages, they also present considerable challenges. 
Within this broader landscape, their role in literature reviews sections has gained particular attention, as AI tools can rapidly scan vast databases, identify relevant studies, summarize findings, and highlight emerging themes~\citep{Quthor_2024}. However, these tools have several drawbacks, such as the tendency to generate fabricated information, known as ``hallucinations,'' which requires rigorous post-hoc human validation for scientific accuracy~\citep{liang2024mapping,bolanos2024artificial,salimi2024llms}. Usability remains a barrier for many existing tools, combining workflow misalignment and reliability issues with the cognitive effort required to craft precise and effective prompts~\citep{bolanos2024artificial,salimi2024llms}.

Although prior work has explored how researchers engage with Generative AI (GenAI) for academic writing~\citep{liao2024llms,bail2024can,gruda2024three}, the practical aspects of this engagement---how researchers structure their LLM-based pipelines, what limitations they encounter, and which strategies they adopt during the literature review process---remain underexplored. In this paper, our goal is to characterize these concrete challenges researchers face when using GenAI for literature review---particularly workflow integration, and trust in model outputs--and to uncover the design bottlenecks that limit current tools and practices. We conducted a user study across different domains, based on the following research questions: 
%~\citep{zimmerman2007research}
% Note: We're not planning for the purpose of this study, so let's scope down to summaries of related works in a formative tool.

\begin{description}
    \item[RQ1.] \textbf{Adoption and Use.} How do researchers integrate GenAI, particularly LLM-based systems, into the process of writing literature reviews? 

    \item[RQ2.] \textbf{Benefits, Challenges and Validation.} What benefits and challenges do researchers encounter when employing GenAI for literature review, and how do researchers verify and ensure the reliability of information retrieved or generated by LLMs? 

    \item[RQ3.] \textbf{Workflow.} What end-to-end workflows emerge for literature review? 

\end{description}

%subsections of main research questions (leave 3 main) -answer in formative study.

\paragraph{\textbf{Main Findings.}} Based on semi-structured interviews, we found that our participants don't trust AI to accurately retrieve or summarize related work due to frequent hallucinations and inaccuracies in generated content. They also move back and forth between verification and generation which leads to a fragmented workflow: researchers often use one tool to find sources, another to summarize or verify content, and a third to draft sections. To maintain accuracy, they rely heavily on verification—cross-checking retrieved papers against original PDFs or DOIs and ensuring that generated findings align with the source material—and on prompt engineering, using instructions such as ``preserve the author’s intent'' or ``not to change my writing style''. 
While these practices are necessary, they are time-consuming, inefficient, and demand considerable effort.

\paragraph{\textbf{Contribution.}} Based on the empirical findings, we propose six design goals for future LLM-based research assistants that emphasize knowledge organization and comparison, citation grounding with stable revisions, preserving author preferences, and guided interaction with transparent rationales. In addition, we propose a high-level framework that operationalizes such goals with human-in-the-loop validation and iterative feedback. We emphasize that GenAI should not function merely as an independent text generator or reviewer, but as an assistant that actively supports researchers in developing their work in a reliable and trustworthy manner.
%through a heterogeneous, citation-grounded publication graph with ontology-guided subgroups, structured comparison views, a per-sentence revision ledger, an Author Profile for stylistic alignment, and a guidance panel for ``why this output?'' explanations and prompt coaching

\section{Related Work}

In addition to the plethora of GenAI applications in research, unified AI frameworks for scientific discovery aim to create autonomous research assistants capable of hypothesis generation, experiment design, and literature synthesis~\citep{reddy2025towards}. These ``AI scientists'' could significantly accelerate cross-domain innovation, provided collaboration continues among AI researchers, domain experts, and philosophers of science~\citep{reddy2025towards}. While fully autonomous discovery remains aspirational, near-term developments promise hybrid systems that augment human scientific reasoning and creativity.

\subsection{Agentic AI}

Agentic AI extends traditional generative systems by endowing them with autonomy, reasoning, and goal-oriented behavior~\citep{gridach2025agentic,huang2024survey}. Unlike simple, single turn chat interactions with generative AI chatbots, agentic AI systems can plan, act, and adapt dynamically within complex environments~\citep{shinn2023reflexion}. They incorporate cognitive mechanisms such as tool calling, memory, reflection, and decision-making, which enable continuous learning and interaction~\citep{shinn2023reflexion, yao2023tree}. %These mechanisms allow agentic system to solve more complex workflows over extended periods of time~\citep{wang2023voyager, park2023generative}. 
In scientific contexts, agents can autonomously conduct literature review, generate hypotheses, create summaries of research bodies, and design experiments, thus streamlining the research process while maintaining interpretability and oversight~\citep{gridach2025agentic,huang2024survey}. This paradigm aligns with the notion of multi-agent systems---autonomous entities that exhibit human-like behavior and collaboration within multi-agent ecosystems~\citep{park2023generative}. The emergence of agentic AI signifies a shift from static computation to adaptive, interactive intelligence that acts as a collaborator rather than a mere tool in scientific inquiry~\citep{gridach2025agentic,huang2024survey}.

\subsection{Generative AI and Agents for Scientific Progress}

LLM-based agents can not only summarize and synthesize literature but also propose new hypotheses and research directions~\citep{wang2024scimon,reddy2025towards}. Nonetheless, studies highlight persistent concerns regarding reliability, bias, and reproducibility across disciplines~\citep{tang2024are,yun2023appraising,meng2024application,saied2024ai,gehrmann2024large,pang2025understanding,mishra2024use}. Hence, while GenAI and agentic systems offer unprecedented support for scientific writing and discovery, human oversight and rigorous verification remain essential to ensure the integrity and credibility of AI-assisted scholarship.

\subsection{Literature Review}

The purpose of a literature review is to develop a thorough understanding of the topic area and to become familiar with existing research before constructing an argument or justification~\citep{arshed2015literature}. A review of relevant literature is an essential component of any academic project, as it provides a firm foundation for advancing knowledge~\citep{webster2002analyzing}. 

A literature review is not merely a summary of prior work but a systematic process of inquiry carried out before and during the writing of a research paper~\citep{machi2009literature, arshed2015literature}. This process involves several stages: selecting and refining a research topic~\citep{machi2009literature, arshed2015literature}, searching for relevant publications using databases, keywords, and citation networks~\citep{machi2009literature, arshed2015literature}, reading and analyzing previous studies to understand their contributions, methods, and gaps, synthesizing insights to identify patterns, themes, and debates, and finally, positioning one’s own research in relation to the existing body of knowledge.

\subsection{Generative AI and Agents for Literature Review}

Recent work demonstrates the increasing use of large language models (LLMs) in literature review and scientific writing~\citep{wang2024scimon,huang2023role,agarwal2024llms,li2024chatcite}. Earlier extractive approaches to automatic related-work generation---such as topic-based summarization~\citep{hoang2010towards}, multi-source pointer-generator networks~\citep{xing2020automatic}, and relation-aware generators~\citep{chen2021capturing}---have evolved into more advanced architectures integrating contrastive and target-aware learning~\citep{chen2022target,ge2021baco}. However, earlier systems struggled with coherence and factuality due to model size and lack of procedural language modeling~\citep{huang2023role}. 

LLMs have since introduced major advances in literature synthesis, being able to recall from large corpora of scientific work~\citep{lewis2020retrieval}. Yet, limitations such as hallucination, outdated information, and opaque reasoning persist~\citep{huang2025survey,gao2023retrieval,li2024dawn}. Retrieval-Augmented Generation (RAG) has emerged as a robust solution, combining LLMs with dynamic knowledge bases to improve factual accuracy and citation grounding~\citep{gao2023retrieval}. Notable implementations, such as ChatCite~\citep{li2024chatcite}, leverage RAG principles for structured citation extraction, re-ranking, and reflective generation workflows. Other research explores LLMs for information retrieval and ranking~\citep{sun2023chatgpt,pradeep2023rankzephyr,hou2024large,wei2021finetuned,crispino2023agent}. Comparative assessment techniques and self-prompting frameworks have further enhanced open-domain question answering and reasoning robustness~\citep{liusie2023llm,li2022self,shi2023prompt}.

This user study examined how researchers integrate LLMs into their literature review workflows, uncovering underexplored challenges in verification, synthesis, and workflow design. While prior work has emphasized efficiency gains and risks such as hallucination, our findings reveal deeper, process-level gaps that shape how trust, accuracy, and conceptual structure are maintained across LLM-assisted review tasks. 

\section{Methods}

We conducted a user study examining how researchers apply LLMs in practice, focusing on their main pipeline processes and the steps they take to verify results. The study was reviewed and approved by the University of Notre Dame’s Institutional Review Board (Protocol 25-06-9369). We compensated each participant with a USD \$25 gift card for their participation. We conducted semi-structured interviews virtually over Zoom that lasted approximately 40 minutes and were recorded for later analysis.

\subsection{Participants}

We recruited eight participants by email who use GenAI for research writing. The group included three Computer Science PhD students and one faculty member, one Biology postdoc and one faculty member, one English PhD student, and one Political Science faculty member. Participants were selected based on their experience with research paper writing. All interviews were conducted under conditions of confidentiality, and participants were informed about the study’s purpose and how their data would be used. Additional participant details are provided in Table~\ref{tab:participants} in the Appendix~\ref{app:A}.

\subsection{Study Procedure}

Each participant took part in an individual remote session conducted via Zoom. At the beginning of the session, we provided a brief introduction outlining the purpose of the study and obtained informed consent (approximately 5 minutes). Participants then engaged in a semi-structured interview lasting between 30 and 45 minutes. During the interview, they were invited to describe or demonstrate their prior interactions with large language model (LLM) systems, discussing their experiences and typical workflows. The interview included questions about participants’ background and additional prompts aligned with the research questions defined in this work (see Appendix~\ref{app:procedures}). 
All sessions were screen-audio-recorded and later transcribed for analysis.

\subsection{Analysis Method}
We analyzed the study data using thematic analysis similar to~\citet{braun2006using}. The sessions were coded based on the transcripts. Codes were iteratively refined and grouped to identify the main themes.

\section{Results}\label{sec:analysis} %study(?)
%%By research questions (and ub research questions) i see paper from professor werner The emmerging use ofr genAI for UX in software development...\\

The insights below, organized by research question, highlight four emerging areas where current practices fall short and where future agent design could substantially improve reliability, reasoning transparency, and reproducibility.

\subsection{RQ1: Adoption and Use}

\textbf{Key Insight 1.1: LLMs are being re-purposed as ``meta-organizers'' that help researchers structure, rather than simply generate, scholarly knowledge.}  
Researchers frequently involve AI in the early stage of research, for example, P6 emphasized using the tool to ``build background knowledge for a specific area". P7 says ``I want to investigate who else is doing XYZ, right? That's related to my own work,'' while P4 uses AI for overall conceptualization, such as figuring out ``what are the most important variables'' in the system they are investigating. This reflects a shift from LLMs as text generators to cognitive scaffolds that assist in framing problems and defining thematic boundaries. They want LLMs to act as information organizers that make conceptual and methodological distinctions explicit---not as summarizers that condense text into generic overviews.
%Across disciplines, participants described using LLMs to \textbf{map conceptual spaces}, \textbf{segment dense topics into frameworks}, and \textbf{design the structure of related work sections}. 

\textbf{Key Insight 1.2: Researchers rely on a diverse ecosystem of LLMs and complementary tools, selecting each for distinct strengths in the literature review process.}  
Participants most often used ChatGPT and Gemini, alongside tools such as Perplexity and Deepseek. These general-purpose LLMs support idea exploration, summarization, and language refinement: P1 uses ChatGPT to ``formulate thoughts'', while P2 and P3 rely on it to ``collect information'' and ``fix language''. Domain-oriented platforms like SciSpace and NotebookLM are preferred when summaries must stay tied to specific PDFs or citations, since general-purpose LLMs can misattribute or hallucinate citations.To mitigate these issues, researchers routinely pair LLMs with verification-focused tools such as LitMaps, Scite.ai, and Google Scholar. As P4 noted, this often means taking ChatGPT’s output and ``copying the title and putting it on Google Scholar or normal Google'' to verify correctness. Tools like NotebookLM are valued, as P8 highlighted, for linking “to the place in the paper” that grounds the generated text.
This multi-tool workflow underscores an unmet opportunity: integrating LLM fluency and synthesis with the citation grounding and reliability of research-specific platforms.

\textbf{Key Insight 1.3: The frequency of LLM use varies widely---from daily integration in research routines to selective, task-specific engagement---reflecting differences in discipline.}  Several participants, particularly from computational and data-intensive fields, reported using LLMs daily as part of their research workflow. P4 described these models as ``a daily toolkit'', and other participants emphasized the use for brainstorming, drafting, and summarizing, with chat sessions often remaining open for days or weeks as evolving workspaces. Others adopt a more occasional, selective approach, engaging LLMs primarily when facing specific writing challenges, such as clarifying ideas, meeting length constraints, or improving readability, for example, P5 said "I don't use [LLMs] daily but just for a couple of essays." This selectiveness stems largely from researchers’ limited trust in LLMs for some complex, intellectual, or fact-intensive tasks, and from varying levels of expertise in using these tools effectively without overreliance.

\textbf{Key Insight 1.4: Researchers primarily seek structured, comparative, and interpretable information from LLMs rather than narrative summaries.}  
Interviewees consistently valued frameworks, taxonomies, and structured hypotheses that clarify relationships across works. They also want systems that can extract hypotheses or results in standardized formats (tables, bullet points, short summaries) to facilitate comparison and synthesis. P4 noted that when ``listing hypotheses, I prefer a table format" and for "summaries of ideas, I prefer bullet points." This concise and straightforward representation helps clarity and verification. The same participant highlighted that “you only need 2 or 3 sentences of the paper or the main results and the implications. So if we could trust AI for that [..]
%, in a way that we can trust, that 
would be really a time saver.”

\subsection{RQ2: Benefits, Challenges and Validation.}

\textbf{Key Insight 2.1: LLMs act as language-level collaborators, improving style and clarity while shaping how researchers express ideas.}  
Rather than serving merely as grammar tools, LLMs function as cognitive amplifiers—supporting researchers at both the language level and the idea level. At the linguistic level, they help users refine grammar, vocabulary, and tone, improving fluency and confidence. P1 said, "I'm not a native speaker, so that's why it helps me to write." At the conceptual level, they assist in articulating and structuring complex arguments, offering feedback akin to a writing coach. P6 described that "it is very helpful to have a [..]
%kind of a mentor or someone, 
coach next to you, [to] tell you what is wrong, what is right." These benefits, however, highlight a growing trade-off between linguistic polish and epistemic fidelity: cleaner text can mask subtle shifts in meaning or evidence attribution, calling for editing systems that integrate clarity with verifiable grounding.

\textbf{Key Insight 2.2: Researchers do not trust LLMs to generate literature reviews autonomously, and even when guided, they remain cautious about the factual reliability and citation integrity of the output.}  
Participants consistently described LLMs as unreliable when producing citations or summarizing scientific sources, reporting that the models frequently invent papers, misattribute findings, or fabricate references. This distrust is pervasive---even among frequent users. P4, who emphasized the need to “always double-check” admitted “I wish I could trust it, but I can’t.” P7 also highlights that "Creating journals out of thin air is a problem 
%That is a problem. 
[..] We all know it." As a result, researchers rarely allow LLMs to autonomously draft related work sections and instead use them only for \textbf{discovery, brainstorming, or organization} after manually verifying all sources.  

This behavior reflects a strong boundary of human oversight: while LLMs are valued for efficiency and linguistic refinement, users insist on direct human verification of evidence and claims before integrating AI-generated text into scholarly writing. Building systems that combine generative capabilities with transparent, citation-grounded verification mechanisms could reduce this distrust and enable safer, more reliable AI-assisted literature review.

\textbf{Key Insight 2.3: The primary bottleneck is not writing quality but maintaining semantic and citation integrity across iterative revisions.}  
Researchers struggle to control semantic drift, where meaning or attribution subtly changes after AI rewriting. Citation positions often shift---moving from one phrase or idea to another unrelated passage or insight---while technical nuances such as argument structure become diluted, and the disciplinary tone of the text is flattened into a neutral, generic style. Several participants noted that LLMs tend to overwrite their unique academic voice---the distinctive way they construct arguments, balance evidence, and express disciplinary identity. As P4 explained, ``I explicitly say not to change my writing style, because ChatGPT will turn it into a kind of generalist text without a soul.'' This highlights the need for verification- and style-aware editing agents capable of detecting and preserving semantic precision, citation alignment, and authorial tone during AI-assisted revisions.

\textbf{Key Insight 2.4: Researchers expect LLMs to surface the core knowledge elements of a paper---its definitions, empirical findings, and author-intended implications---rather than independently infer or generate new content.}  
Many emphasized the need for AI to distinguish between factual information, the author’s interpretation (e.g., ``according to the author''), and AI-generated information. As P4 noted, ``If you don’t say, 'according to the author' it is gonna make its own interpretation, and when you are reading other people’s works you don’t know that. You want what the person actually wrote.'' Current tools lack this fine-grained control and often blur factual claims with inferred meaning. %Designing \textbf{citation-grounded extraction} that explicitly attributes claims to textual evidence could bridge this gap.

\textbf{Key Insight 2.5: Finding contextually relevant papers remains one of the most significant challenges researchers face when using LLMs or search tools for literature discovery.} Participants emphasized that identifying papers is difficult not because results are off-topic (not related), but because few are genuinely relevant to their specific research focus. They described relevance as having two dimensions: (1) Contextual relevance, which involves recognizing the author’s expertise, publication venue, and disciplinary importance; and (2) Conceptual alignment, referring to works that share a similar rationale, methodological approach, or research question. As P6 explained, ``I’m trying to see more the rationale [..] like why and where this paper fits in the research area. It’s hard to say this paper is very relevant; maybe, why is that? Sometimes it’s just because Google Scholar knows who the researcher is or that it already has many citations''. However, they also noted that knowing an author’s reputation or that a paper appeared in a prestigious venue provides ``a much better sense of relevance.'' Participants therefore stressed that current LLM recommendations lack this contextual understanding, often treating all topical overlaps as equally important. They expressed a need for systems that combine topical relevance with contextual cues (e.g., author networks, venues, and exploration of adjacent ideas) to better surface key literature.
%However, P6 also  highlights, “If you can see the authors, or know that this paper was published at ICML versus a random conference, you get a much better sense of relevance.” 
\subsection{RQ3: Workflow.}

\textbf{Key Insight 3.1: Authors avoid accumulation of content across sessions and highlight the lack of traceable verification.}  
Researchers often split their work into multiple chats to prevent context pollution, as reported by P3: ``I use multiple chats'' because ``[I] don't want [..] the previous paragraphs that I changed, like, polluting the context for new ones'' which ironically creates verification silos. There is no persistent record of which outputs were verified, how sources were confirmed, or whether AI-suggested content survived later edits, as P8 reported: ``[LLM] misconstrues the meaning that I intended in the original [paragraph].'' The absence of a traceable verification hinders reproducibility and accountability. This highlights a critical design need for a persistent verification memory that tracks, annotates, and surfaces the provenance and validation status of AI-assisted contributions over time.

\begin{figure*}[h]
  \centering
  \includegraphics[width=0.7\linewidth]{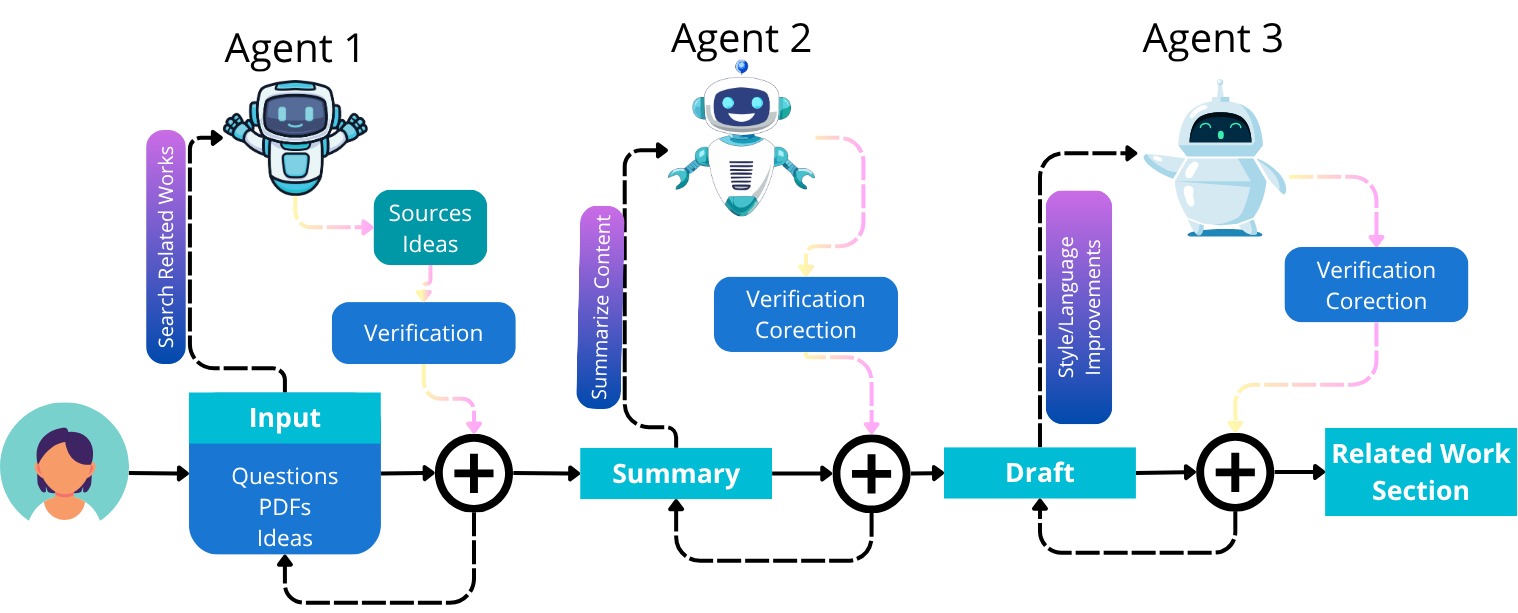}
  \caption{Researcher's workflow when doing literature review with GenAI based on the interviews.}
   \label{fig:workflow}
\end{figure*}

\textbf{Key Insight 3.2: Researchers follow a hybrid, cyclical pipeline for conducting literature reviews. }  
As illustrated in Figure~\ref{fig:workflow}, the pipeline typically begins with researchers defining initial prompts or identifying seed papers, after which the LLM retrieves potentially relevant works. These suggestions then undergo an initial verification stage, where researchers manually confirm that the papers align with the intended research scope. Next, the LLM generates summaries or key points for the selected papers, which again require manual review to ensure factual accuracy and faithful representation of the authors’ claims. In later stages, researchers use the LLM to draft or refine text; while this improves coherence and argumentation, it demands another round of verification to ensure that meaning is preserved and no unverified interpretations are introduced. Through iterative cycles of retrieval, summarization, drafting, and verification, the related work section emerges as a human–AI co-authored product. However, each stage currently operates in isolation---often involving multiple LLMs---requiring researchers to manually track sources, versions, and evidence. This fragmentation highlights the need for integrated, verification-aware systems that unify the workflow while preserving provenance, accuracy, and conceptual structure across iterations.

%As illustrated in Figure~\ref{fig:workflow}, the pipeline typically begins with researchers defining initial prompt ideas or identifying seed papers, after which the LLM retrieves potentially relevant works. These suggestions then undergo an initial verification stage, where researchers manually confirm that the papers align with the intended research scope and topics. Next, the LLM generates summaries or key points for the selected papers, which again require manual review to ensure factual accuracy and faithful representation of the authors’ claims. In later stages, researchers use the LLM to draft or refine text. While this step improves coherence and argumentation, it demands another round of verification to confirm that the LLM has not altered the original meaning or introduced unverified interpretations. Through multiple iterations of retrieval, summarization, drafting, and verification, the related work section gradually takes shape as a human–AI co-authored product. However, each stage currently operates in isolation, requiring researchers to manually track sources, versions, and evidence. Also, this often requires the use of multiple LLMs, as users prefer different LLMs for different tasks. This fragmentation underscores the need for integrated, verification-aware systems that connect all steps into a continuous workflow---preserving provenance, accuracy, and conceptual structure across iterations.

\section{Design Goals}%implications (?)

Based on the key insights from our study, we outline five design goals to guide the creation of future LLM-based systems for literature review. %These goals focus on improving how researchers can organize knowledge, verify information, and build trust in AI-assisted writing.

\begin{itemize}
    \item \textbf{DG1. Domain-Based Knowledge Organization (KI 1.1 \& 1.4):}
    Researchers use LLMs to organize ideas, but current systems lack an understanding of how knowledge is structured within each discipline. Future tools should help users build and refine conceptual maps or frameworks that reflect how works are connected within a specific domain---showing relationships between studies, methods, and themes. This can be achieved through graph-based visualizations and domain-specific representations that align with established scientific taxonomies and citation networks.
    
    \item \textbf{DG2. Structured Comparison and Relevance Filtering (KI 1.4, 2.4 \& 2.5):}
    Instead of producing long, unstructured summaries, LLMs should generate concise comparative outputs that highlight the main methods, results, and insights of each work and how they relate to the researcher’s main idea. Systems should also incorporate relevance-aware retrieval by considering factors such as citation context, publication venue, and author expertise, ensuring that suggested papers are not only related by the same topic (supported by systems such as Litmaps, and Scite.ai) but are  relevant to the intended research focus.
    
    \item \textbf{DG3. Citation-Grounded Summaries and Human-in-the-Loop Verification (KI 2.1, 2.3 \& 3.2):}  
    A major challenge is maintaining consistent meaning and citation accuracy when revising text with AI. Systems should not only display which source supports each idea but also how that idea is represented in the original context. Each statement should be explicitly linked to its originating passage, such as a paragraph, figure, or result section, allowing users to inspect whether the AI’s phrasing preserves the author’s intent, emphasis, and evidentiary function of the citation.
   Also, the system should track revisions across sessions, changes in relation between the AI’s paraphrase and the original claim, and alert users when interpretations drift or when citations are repurposed beyond their intended scope. Finally, the system should maintain version history to support transparent, human-in-the-loop verification over time.

    %purpose of the citation keeps the same (purpose tracking)
    
    \item  \textbf{DG4. Author-Aware Expression and Domain Adaptation (KI 2.1 \& 2.4):} Researchers value outputs that reflect their own writing style and disciplinary tone. Systems should distinguish between author-level adaptation---anchoring phrasing and structure to the researcher’s notes, prior drafts, and stylistic preferences---and domain-level adaptation, which adjusts vocabulary, framing, and argumentative conventions to fit the norms of a given field.
    This helps preserve authenticity and consistency in AI-assisted writing.
    
    \item \textbf{DG5. Guided Interaction and Transparency (KI 2.2, 3.1 \& 3.2):}  
    Many users rely on long or complex prompts to obtain accurate results, which often makes the process tiring and inconsistent. To address this, LLM tools should offer interactive guidance that helps users formulate clearer, more efficient queries. By explaining why certain outputs were generated and suggesting better ways to ask questions, systems can improve both usability and verification. The objective is not to train users in advanced prompting techniques, but to design systems capable of handling natural, non-expert queries while maintaining transparency on how responses are derived.
    \item \textbf{DG6. Validation (KI 2.2, 3.1 \& 3.2):}
    The framework should improve verification and trust by positioning the LLM as a “judge” within the review pipeline---assessing factual consistency, citation validity, and semantic stability at each step.
\end{itemize}

These goals outline how new LLM-based systems can go from basic text generation toward \textbf{trustworthy, structured, and verifiable collaboration} between humans and AI. 

\section{Proposed Framework for LLM-Assisted Literature Review}

\begin{figure*}[h]
  \centering
  \includegraphics[width=0.93\linewidth]{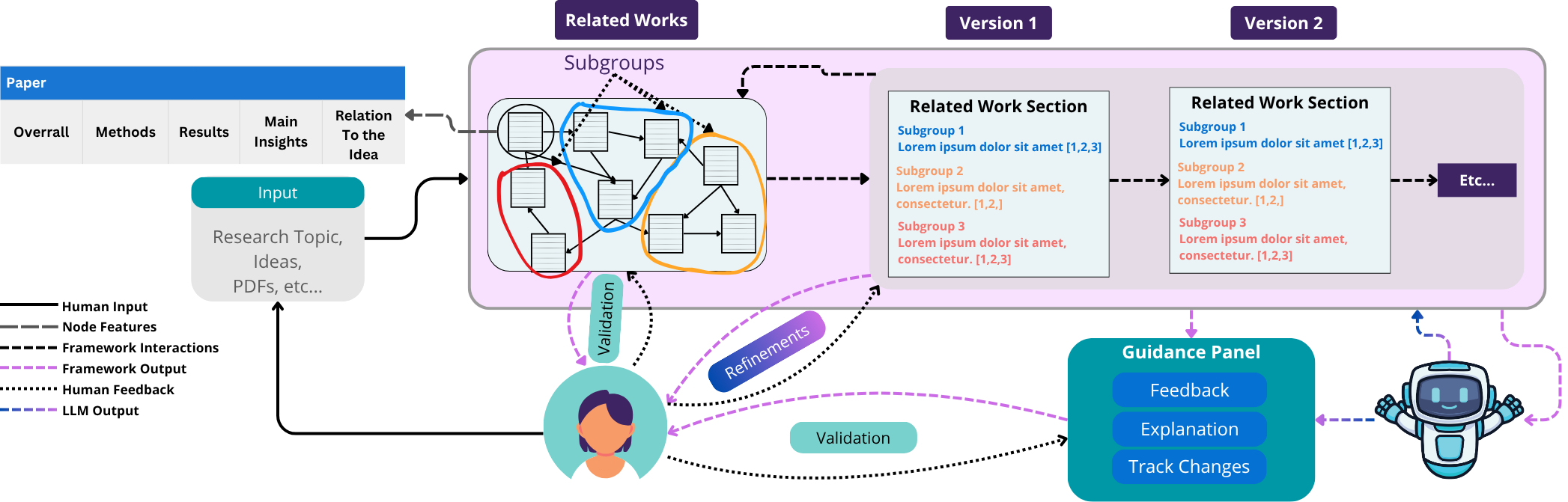}
  \caption{End-to-end framework for an ideal literature-review system implementing design goals DG1–DG6. The system ingests user inputs and reference papers, constructs a heterogeneous knowledge graph with thematic subgroups, incorporates researcher preferences into generation, produces citation-anchored drafts with a revision ledger for semantic stability, and offers a guidance panel for transparency, feedback, and validation.}
   \label{fig:design}
\end{figure*}

We propose an overall framework for an \textit{ideal} literature review system that operationalizes the design goals DG1–DG6 (Figure~\ref{fig:design}). The system is built around a domain-aware knowledge graph that integrates scholarly papers, researcher inputs, and citation context. Our goal is not to detail implementation but to outline the overall workflow and the conceptual architecture that could guide future systems.

\subsection{Core Components}
\begin{itemize}
    \item \textbf{Paper Graph (\(G\))}: A heterogeneous, citation-grounded graph where nodes represent papers and edges encode relations such as citation, methodological similarity, dataset overlap, and conceptual affinity.
    \item \textbf{Node Card}: A structured representation of papers containing (i) overall metadata (title, venue, authors), (ii) methods, (iii) main results, (iv) key insights, and (v) explicit links to research idea or problem.
    \item \textbf{Subgroups (Themes)}: Labeled subgraphs that cluster papers by their conceptual or methodological role in the review (e.g., representations, learning methods, evaluation, applications).
    \item \textbf{Author Profile}: A personalized record of the researcher’s preferred writing style, domain terminology, and tone, derived from personal notes or previous publications.
    \item \textbf{Revision Ledger}: A traceable record of edits and changes to ensure consistency of meaning and citation integrity across AI-assisted revisions.
    \item \textbf{Guidance Panel}: A companion module providing interactive explanations, transparency, validation, and prompting feedback, allowing users to understand and refine interactions with the system.
\end{itemize}

\subsection{System Overview}

\paragraph{Feature A: Ingestion and Search (DG1 and DG2).}
The system begins by collecting inputs such as ideas, research questions, preferences, reference papers (as PDFs), and user notes. These inputs form the foundation for a context model that captures the user’s research scope, terminology, and disciplinary focus. It searches for relevant related works (DG2) that ranked in relevance using a composite relevance score combining semantic similarity, citation importance, venue quality, and author expertise. Then the system extracts metadata, sections, and citation contexts, while inferring the research domain through keyword and methodology analysis (DG1). Domain-specific ontologies (e.g., chemistry, HCI) are loaded to guide subsequent structuring (DG1).

\paragraph{Feature B: Graph Construction and Thematic Structuring.}
A domain-aware graph \(G\) is built by defining multiple edge types: citation links, methodological similarity, dataset overlap, and conceptual relationships (DG1). Subgroups are detected through community clustering and ontology-guided labeling, yielding thematic clusters such as ``Representations,'' ``Learning Methods,'' or ``Evaluation and Metrics.'' The system generates structured comparison views---method tables, result tables, and insight maps---where every entry links back to the source paragraph (DG2). This enables concise, comparative, and transparent synthesis. In general, each node card summarizes a paper’s main contributions and its relation to the target idea, while edges reveal methodological or conceptual overlap. The researcher can pin or unpin nodes (papers).

\paragraph{Feature C: Author Curation and Preference Binding.}
The researcher filters the graph \(G\) to create a curated subgraph \(G^{*}\) (DG1), selecting relevant papers and excluding marginal ones (DG2). The system then aggregates the Author Profile in the generation pipeline, ensuring that future text generation aligns with a researcher’s writing style and preferred terminology (DG4).

\paragraph{Feature D: Citation-Grounded Drafting and Stable Revisions.}
Based on \(G^{*}\), the system proposes a structured outline that mirrors the thematic subgroups. The LLM generates a draft Related Work section that is concise, comparative, and fully citation-anchored, i.e. each sentence is explicitly linked to one or more sources (DG4). The Revision Ledger monitors all changes, flags semantic drifts, and preserves citation integrity over time (DG3).

\paragraph{Feature E: Guidance and Transparency.}
The Guidance Panel explains why specific outputs were generated, which nodes or author preferences influenced them, suggests clearer ways to formulate questions, and surfaces validations to the author (DG5 \& DG6). Over the different interactions of refinement, the system tracks changes in affirmations and citations, and returns to the researcher to validate or not those modifications in the generated text (DG6). Beyond user feedback, it embeds an automated validation layer where the LLM functions as a ``judge,'' cross-checking citations and summaries against the retrieved papers, verifying semantic veracity between summaries and sources, and flagging inconsistencies in claims or attribution (DG6). The Guidance Panel may, for example, alert the user that ``most studies rely on certain metrics'' and suggest including a specific data for completeness. It also should detect potential biases, missing baselines, or imbalanced citation coverage, suggesting relevant additions to maintain scholarly rigor. %These proactive checks emulate human verification but automate it, allowing researchers to confirm flagged issues directly within the interface—turning validation into a continuous, system-driven process rather than a manual, post-hoc task.

%\subsection{Example Domain: Chemistry}
%To illustrate, consider the target idea \textit{``How molecular representations and learning methods affect potency prediction in drug discovery.''}  
%The system would construct thematic subgroups (S) as follows:
%\begin{itemize}
    %\item \textbf{S1: Representations} -- including ECFP/Morgan fingerprints, graph neural networks, 3D embeddings, and SMILES-based models.
    %\item \textbf{S2: Learning Methods} -- comparing random forests, GNN variants, multitask learning, and pretraining strategies.
    %\item \textbf{S3: Evaluation and Data} -- aggregating datasets (e.g., ChEMBL), metrics, such as RMSE, MAE or SERA~\cite{Ribeiro2020}, and evaluation protocols.
   % \item \textbf{S4: Applications and Findings} -- linking case studies such as EGFR potency prediction and virtual screening frameworks.
%\end{itemize}

%In general, each node card summarizes a paper’s main contributions and its relation to the target idea, while edges reveal methodological or conceptual overlap. The researcher can pin or unpin nodes, and the LLM produces a concise related-work section that contrasts key approaches while maintaining proper citation alignment. The Guidance Panel may, for example, alert the user that ``most studies rely on certain metrics'' and suggest including a specific data for completeness. It also should detect potential biases, missing baselines, or imbalanced citation coverage, suggesting relevant additions to maintain scholarly rigor.

\section{Conclusion and Future Work}

This paper analyzes how researchers incorporate LLMs into literature reviews and, from observed practices and challenges, distills six design goals for trustworthy and verifiable human–AI collaboration. We propose a domain-aware, citation-grounded framework that structures prior work through ontology-guided knowledge graphs, favors structured comparison over narrative summarization, and anchors all generated claims to verified source passages. Interactive controls expose system reasoning and help users ensure accuracy and traceability.

Our framework positions the LLM as a collaborative evaluator that cross-checks evidence, flags unsupported statements, and prompts users to confirm or correct claims. We plan to prototype and evaluate this approach with researchers during related-work drafting and revision, assessing faithfulness, edit stability, and user trust, and exploring extensions involving community ontologies, venue and author signals, and cross-domain generalization. %Finally, we intend to implement a persistent verification memory that accumulates validation status over time and provides transparent, sentence-level audit trails linking generated text to its source evidence.

By treating the literature review as a process of graph construction and curation rather than one-shot text generation, the framework reframes LLM-based literature review as an agentic, collaborative experience---similar to vibe coding---where researchers engage in conversational interaction with an agent that navigates and refines the evolving knowledge graph and the final product---the related work section.

\subsection{Limitations and Ethical Considerations}
Given our study’s focus on early GenAI adopters, broader sampling may reveal additional needs—particularly in disciplines with distinct evidentiary norms. Proposed relevance signals (e.g., venue or author reputation) could also reinforce existing hierarchies; mitigation will require transparent weighting, user controls, and regular audits.

The integration of AI into scientific ecosystems unfolds within entrenched systems of inequality. Prior work shows how advantages accumulate over time \citep{merton1968matthew}, women’s contributions are systematically undervalued \citep{rossiter1993matilda}, and institutional prestige shapes whose research gains visibility \citep{kozlowski2024howard}. These dynamics intersect with geographic and linguistic biases, producing center–periphery structures that concentrate authority in elite, English-speaking institutions \citep{sugimoto2018lariviere, lariviere2015oligarchy}. Because models learn from these biased systems, they risk reproducing them at scale; algorithmic bias may transform human gatekeeping into automated exclusion, amplifying systemic inequalities and narrowing who benefits from AI-driven discovery.

\section{Acknowledgments}

This research was supported in part by the Notre Dame-IBM Technology Ethics Lab. Brenda Nogueira was partially supported by a fellowship of the Scientific Artificial Intelligence (SAI) Initiative at the University of Notre Dame.

\appendix
\setcounter{secnumdepth}{2}
\begin{strip}

\centering
\resizebox{0.95\linewidth}{!}{%
{\LARGE
\begin{tabular}{p{2.5cm} c c c c}
\hline
\textbf{Participant ID} & \textbf{Occupation} & \textbf{Research Area} &
\textbf{Years of LLM Experience} & \textbf{LLM Tools Previously Used} \\
\hline
P1 & PhD Student & Computer Science & 1-2 Years & ChatGPT, Claude \\
P2 & PhD Student & Computer Science & 1-2 Years & ChatGPT \\
P3 & PhD Student & Computer Science & 1-1.5 Years & ChatGPT, Gemini \\
P4 & Postdoc & Biology & $>$2 Years & ChatGPT, LitMaps, Perplexity, SciSpace, Gemini \\
P5 & PhD Student & English & 2 Years & ChatGPT \\
P6 & Faculty & Computer Science & 2-3 Years & ChatGPT \\
P7 & Faculty & Political Science & 1-1.5 Years & ChatGPT \\
P8 & Faculty & Biology & 2-3 Years & ChatGPT, Gemini, Claude, NotebookLM, Humata, DeepSeek, Thesify \\
\hline
\end{tabular}
}%
}
\captionsetup{type=table}
\captionof{table}{Demographic information of participants in the user study.}
\label{tab:participants}

    \small
    \centering
    \resizebox{0.95\linewidth}{!}{%
    \begin{tabular}{p{0.2\linewidth}|p{0.05\linewidth}|p{0.75\linewidth}}
    \toprule
    \textbf{Phase} & \textbf{RQ\#} & \textbf{Question} \\
    \midrule
    
    % ---------------- BACKGROUND ----------------
    \multirow{2}{*}{Background} 
        & RQ0 & What is your main field of study or work? \\  
        &     & What is your current position in your field of study? \\
    \hline
    
    % ---------------- ADOPTION AND USE ----------------
    \multirow{16}{*}{Adoption and Use} 
        & RQ1 & At what stage of the writing process do you usually involve AI tools? \\
        &     & \hspace{0.5cm}\textit{Is it during early exploration, drafting, revising, or another stage?} \\
        &     & How long have you used AI tools to assist with research writing? \\
        &     & How frequently do you use these AI tools in your research writing? \\
        &     & Which LLMs or AI tools do you typically use? \\
        &     & The last time you started writing a related works section, what was the first thing you asked the AI agent (e.g., ChatGPT)? \\
        &     & What kind of information do you usually provide to the agent at the beginning, and in what format? \\
        &     & What kind of output are you expecting from the agent? \\
        &     & \hspace{0.5cm}\textit{Individual sentences, paragraphs, bullet summaries, or full sections?} \\

        &     & How do you typically give feedback to the agent during interaction? \\
        &     & \hspace{0.5cm}\textit{Do you ask it to rewrite, elaborate, simplify, or include more technical detail?} \\
        &     & Do you interact with the agent in a single long session or across multiple sessions? \\
        &     & How do you usually perform related work without using LLMs? \\
        &     & \hspace{0.5cm}\textit{What steps or resources do you rely on in that process?} \\
        &     & Have you found specific strategies or prompt styles that work particularly well for this task? \\
        &     & How important is AI assistance for the related works section compared to other parts of your paper? \\
    \hline
    
    % ---------------- BENEFITS, CHALLENGES AND VALIDATION ----------------
    \multirow{11}{*}{\parbox{3cm}{Benefits, Challenges, and Validation}}

        & RQ2 & What are the most helpful aspects of using AI in your literature review workflow? \\
        &     & What challenges or limitations have you experienced when using AI tools? \\
        &     & What types of AI assistance would you like to have, or what gaps do you notice in current tools? \\
        
        &     & \hspace{0.5cm}\textit{How do you use AI for other sections of your research paper?} \\
        &     & Do you trust the information directly, or do you verify it in any way? \\
        &     & What does “changing for your own style” mean to you? \\
        &     & \hspace{0.5cm}\textit{Does it mean improving readability or aligning with your community’s conventions?}
        \\
         &     & How do you know when the AI has hallucinated or provided incorrect information? \\
          &  & If you use the agent for citation suggestions, how do you handle the references it provides? \\
        &     & \hspace{0.5cm} Do you verify the sources? If so, how? \\
         &     & What strategies do you use to ensure that generated text and citations are accurate and aligned with your argument? \\
    \hline
    
    % ---------------- WORKFLOW DESIGN ----------------
    \multirow{4}{*}{Workflow} 
        &   RQ3  & How do you perceive the role of GenAI — as a co-writer, reviewer, or analytical assistant? \\
        &  & \hspace{0.5cm}\textit{How does your interaction evolve as the related works section develops?} \\
        
        &     & How do you incorporate human feedback or verification during the process? \\
        &     & What does your typical end-to-end workflow with AI look like — from initial query to final revision? \\
    
    \bottomrule
    \end{tabular}}
    \captionsetup{type=table}
    \captionof{table}{Interview guide aligned with research questions (RQ0–RQ3) on LLM-assisted literature review practices.}
    \label{tab:interview_script_rq}
\end{strip}

\section{Appendix}
\subsection{Participants}
\label{app:A}

We provide more details about the participants in Table~\ref{tab:participants}.

\subsection{Procedures}
\label{app:procedures}

The main questions asked during the interview are presented in Table~\ref{tab:interview_script_rq}.

% \subsection{Protocol}\label{app:protocol}

%
%The IRB protocol used in this work starts in Figure~\ref{fig:protocol}.
%\clearpage
%\setcounter{figure}{0}
%\includepdf[pages={3-23,25-}, scale=0.8,pagecommand={
 %\begingroup
    %\setcounter{figure}{1}%
     % \refstepcounter{figure}
     % \captionof{figure}{Selected pages from the IRB Protocol Submission Form.}
      %\label{fig:protocol}
    %\endgroup
%  },
%  pagebox=cropbox
%]{Protocol Submission FormIRB_25-06-9369.pdf}

\clearpage
\bibliography{aaai2026}

@misc{Quthor_2024, title={Understanding the necessity of AI in academic writing}, url={https://quickcreator.io/blog/necessity-of-ai-in-academic-writing/}, journal={Elevate Your Online Presence with AI-Driven, SEO-Enhanced Blogs & Landing Pages}, publisher={blog}, author={Quthor}, year={2024}, month={Sep}}

@misc{Kannan_2024, title={How much research is being written by large language models?}, url={https://hai.stanford.edu/news/how-much-research-being-written-large-language-models}, journal={Stanford HAI}, author={Kannan, Prabha}, year={2024}, month={May}}

@article{bolanos2024artificial,
  title={Artificial intelligence for literature reviews: Opportunities and challenges},
  author={Bolanos, Francisco and Salatino, Angelo and Osborne, Francesco and Motta, Enrico},
  journal={Artificial Intelligence Review},
  volume={57},
  number={10},
  pages={259},
  year={2024},
  publisher={Springer}
}

@article{si2024can,
  title={Can llms generate novel research ideas},
  author={Si, Chenglei and Yang, Diyi and Hashimoto, Tatsunori},
  journal={A large-scale human study with},
  volume={100},
  year={2024}
}

@article{liang2024mapping,
  title={Mapping the increasing use of LLMs in scientific papers},
  author={Liang, Weixin and Zhang, Yaohui and Wu, Zhengxuan and Lepp, Haley and Ji, Wenlong and Zhao, Xuandong and Cao, Hancheng and Liu, Sheng and He, Siyu and Huang, Zhi and others},
  journal={arXiv preprint arXiv:2404.01268},
  year={2024}
}

@article{salimi2024llms,
  title={LLMs and Academic Writing in Practice: Exploring Participants’ Utilization of GPT during an AI-Assisted Course on Writing Research Papers},
  author={Salimi, Esmaeel Ali and Hajinia, Mahdi},
  year={2024}
}

@article{braun2006using,
  title={Using thematic analysis in psychology},
  author={Braun, Virginia and Clarke, Victoria},
  journal={Qualitative research in psychology},
  volume={3},
  number={2},
  pages={77--101},
  year={2006},
  publisher={Taylor \& Francis}
}

@article{liao2024llms,
  title={LLMs as Research Tools: A Large Scale Survey of Researchers' Usage and Perceptions},
  author={Liao, Zhehui and Antoniak, Maria and Cheong, Inyoung and Cheng, Evie Yu-Yen and Lee, Ai-Heng and Lo, Kyle and Chang, Joseph Chee and Zhang, Amy X},
  journal={arXiv preprint arXiv:2411.05025},
  year={2024}
}

@article{tang2024are,
  title        = {Are LLMs Good Literature Review Writers? Evaluating the Literature Review Writing Ability of Large Language Models},
  author       = {Tang, Xuemei and Duan, Xufeng and Cai, Zhenguang G.},
  year         = {2024},
  eprint       = {2412.13612},
  archivePrefix= {arXiv},
  primaryClass = {cs.CL},
  journal      = {arXiv preprint arXiv:2412.13612},
  url          = {https://arxiv.org/abs/2412.13612}
}

@article{li2024chatcite,
  title        = {ChatCite: LLM Agent with Human Workflow Guidance for Comparative Literature Summary},
  author       = {Li, Yutong and Chen, Lu and Liu, Aiwei and others},
  year         = {2024},
  eprint       = {2403.02574},
  archivePrefix= {arXiv},
  primaryClass = {cs.CL},
  journal      = {arXiv preprint arXiv:2403.02574},
  url          = {https://arxiv.org/abs/2403.02574}
}

@misc{Agarwal2024LLMs,
  title        = {LLMs for Literature Review: Are we there yet?},
  author       = {Shubham Agarwal and Gaurav Sahu and Abhay Puri and others},
  year         = {2024},
  month        = {December},
  eprint       = {2412.15249},
  archivePrefix= {arXiv},
  primaryClass = {cs.IR},
  doi          = {10.48550/arxiv.2412.15249},
  note         = {arXiv preprint},
}

@misc{Pang2025understanding,
  title        = {Understanding the LLM-ification of CHI: Unpacking the Impact of LLMs at CHI through a Systematic Literature Review},
  author       = {Rock Yuren Pang and Hope Schroeder and Kendal Smith and others},
  year         = {2025},
  month        = {January},
  eprint       = {2501.12557},
  archivePrefix= {arXiv},
  primaryClass = {cs.HC},
  doi          = {10.48550/arxiv.2501.12557},
  note         = {arXiv preprint},
}

@inproceedings{saied2024ai,
  title        = {AI in Literature Reviews: a survey of current and emerging methods},
  author       = {Mohamed Saied and Norehan Mokhtar and A. Badr and others},
  booktitle    = {Proc. MIUCC 2024},
  year         = {2024},
  month        = {November},
  doi          = {10.1109/miucc62295.2024.10783597},
}

@misc{Yun2023appraising,
  title        = {Appraising the Potential Uses and Harms of LLMs for Medical Systematic Reviews},
  author       = {Hye Sun Yun and Iain J. Marshall and Thomas A. Trikalinos and others},
  year         = {2023},
  month        = {May},
  eprint       = {2305.11828},
  archivePrefix= {arXiv},
  primaryClass = {cs.CY},
  doi          = {10.48550/arXiv.2305.11828},
  note         = {arXiv preprint},
}

@article{meng2024application,
  title        = {The Application of Large Language Models in Medicine: A Scoping Review},
  author       = {Xiangbin Meng and Xiangyu Yan and Kuo Zhang and others},
  journal      = {iScience},
  year         = {2024},
  month        = {March},
  doi          = {10.1016/j.isci.2024.109713},
}

@article{mishra2024use,
  title        = {Use of large language models as artificial intelligence tools in academic research and publishing among global clinical researchers},
  author       = {Tanisha Mishra and Edward Sutanto and Rini Rossanti and others},
  journal      = {Dental Science Reports},
  year         = {2024},
  month        = {December},
  doi          = {10.1038/s41598-024-81370-6},
  note         = {Study reports 54.2\% outlook; check journal details},
}

@misc{Gehrmann2024large,
  title        = {Large Language Models for Literature Reviews - an Exemplary Comparison of LLM-based Approaches with Manual Methods},
  author       = {Julia Gehrmann and Lars Quakulinski and Oya Beyan},
  year         = {2024},
  eprint       = {10852447},
  archivePrefix= {IEEE},
  doi          = {10.1109/fllm63129.2024.10852447},
  note         = {Proceedings paper},
}

@article{bail2024can,
  title={Can Generative AI improve social science?},
  author={Bail, Christopher A},
  journal={Proceedings of the National Academy of Sciences},
  volume={121},
  number={21},
  pages={e2314021121},
  year={2024},
  publisher={National Academy of Sciences}
}

@article{gruda2024three,
  title={Three ways ChatGPT helps me in my academic writing},
  author={Gruda, Dritjon},
  journal={Nature},
  volume={10},
  year={2024},
  publisher={Nature Publishing Group}
}

@inproceedings{reddy2025towards,
  title={Towards scientific discovery with generative ai: Progress, opportunities, and challenges},
  author={Reddy, Chandan K and Shojaee, Parshin},
  booktitle={Proceedings of the AAAI Conference on Artificial Intelligence},
  volume={39},
  number={27},
  pages={28601--28609},
  year={2025}
}

@inproceedings{wang2024scimon,
  title={Scimon: Scientific inspiration machines optimized for novelty},
  author={Wang, Qingyun and Downey, Doug and Ji, Heng and Hope, Tom},
  booktitle={Proceedings of the 62nd Annual Meeting of the Association for Computational Linguistics (Volume 1: Long Papers)},
  pages={279--299},
  year={2024}
}

@inproceedings{hoang2010towards,
  title={Towards automated related work summarization},
  author={Hoang, Cong Duy Vu and Kan, Min-Yen},
  booktitle={Coling 2010: Posters},
  pages={427--435},
  year={2010}
}

@inproceedings{xing2020automatic,
  title={Automatic generation of citation texts in scholarly papers: A pilot study},
  author={Xing, Xinyu and Fan, Xiaosheng and Wan, Xiaojun},
  booktitle={Proceedings of the 58th Annual Meeting of the Association for Computational Linguistics},
  pages={6181--6190},
  year={2020}
}

@inproceedings{ge2021baco,
  title={BACO: A background knowledge-and content-based framework for citing sentence generation},
  author={Ge, Yubin and Dinh, Ly and Liu, Xiaofeng and Su, Jinsong and Lu, Ziyao and Wang, Ante and Diesner, Jana},
  booktitle={Proceedings of the 59th Annual Meeting of the Association for Computational Linguistics and the 11th International Joint Conference on Natural Language Processing (Volume 1: Long Papers)},
  pages={1466--1478},
  year={2021}
}

@inproceedings{chen2021capturing,
  title={Capturing relations between scientific papers: An abstractive model for related work section generation},
  author={Chen, Xiuying and Alamro, Hind and Li, Mingzhe and Gao, Shen and Zhang, Xiangliang and Zhao, Dongyan and Yan, Rui},
  year={2021},
  organization={Association for Computational Linguistics (ACL)}
}

@inproceedings{chen2022target,
  title={Target-aware abstractive related work generation with contrastive learning},
  author={Chen, Xiuying and Alamro, Hind and Li, Mingzhe and Gao, Shen and Yan, Rui and Gao, Xin and Zhang, Xiangliang},
  booktitle={Proceedings of the 45th international ACM SIGIR conference on research and development in information retrieval},
  pages={373--383},
  year={2022}
}

@article{huang2023role,
  title={The role of ChatGPT in scientific communication: writing better scientific review articles},
  author={Huang, Jingshan and Tan, Ming},
  journal={American journal of cancer research},
  volume={13},
  number={4},
  pages={1148},
  year={2023}
}

@article{gridach2025agentic,
  title={Agentic Artificial Intelligence for Scientific Discovery},
  author={Gridach, Mourad and Shaban, Ameer and Kocyigit, Naci and et al.},
  journal={arXiv preprint arXiv:2503.08979},
  year={2025},
  url={https://arxiv.org/abs/2503.08979}
}

@article{huang2024survey,
  title={A Survey on Agentic Large Language Models: Architectures, Applications, and Open Problems},
  author={Huang, Yuzhong and Peng, Baolin and Shen, Tianyi and Zhang, Yutong and Gao, Jianfeng},
  journal={arXiv preprint arXiv:2407.01502},
  year={2024},
  url={https://arxiv.org/abs/2407.01502}
}

@article{park2023generative,
  title={Generative Agents: Interactive Simulacra of Human Behavior},
  author={Park, Joon Sung and O'Brien, Joseph C. and Cai, Carrie J. and Morris, Meredith Ringel and Liang, Percy and Bernstein, Michael S.},
  journal={arXiv preprint arXiv:2304.03442},
  year={2023},
  url={https://arxiv.org/abs/2304.03442}
}

@article{shinn2023reflexion,
  title={Reflexion: Language Agents with Verbal Reinforcement Learning},
  author={Shinn, Noah and Labash, Zachary and Gopinath, Ashwin and Madaan, Aman and Liu, Percy and Chen, Yisen and Zettlemoyer, Luke and others},
  journal={arXiv preprint arXiv:2303.11366},
  year={2023},
  url={https://arxiv.org/abs/2303.11366}
}

@article{lewis2020retrieval,
  title={Retrieval-augmented generation for knowledge-intensive nlp tasks},
  author={Lewis, Patrick and Perez, Ethan and Piktus, Aleksandra and Petroni, Fabio and Karpukhin, Vladimir and Goyal, Naman and K{\"u}ttler, Heinrich and Lewis, Mike and Yih, Wen-tau and Rockt{\"a}schel, Tim and others},
  journal={Advances in neural information processing systems},
  volume={33},
  pages={9459--9474},
  year={2020}
}

@article{huang2025survey,
  title={A survey on hallucination in large language models: Principles, taxonomy, challenges, and open questions},
  author={Huang, Lei and Yu, Weijiang and Ma, Weitao and Zhong, Weihong and Feng, Zhangyin and Wang, Haotian and Chen, Qianglong and Peng, Weihua and Feng, Xiaocheng and Qin, Bing and others},
  journal={ACM Transactions on Information Systems},
  volume={43},
  number={2},
  pages={1--55},
  year={2025},
  publisher={ACM New York, NY}
}

@article{gao2023retrieval,
  title={Retrieval-augmented generation for large language models: A survey},
  author={Gao, Yunfan and Xiong, Yun and Gao, Xinyu and Jia, Kangxiang and Pan, Jinliu and Bi, Yuxi and Dai, Yixin and Sun, Jiawei and Wang, Haofen and Wang, Haofen},
  journal={arXiv preprint arXiv:2312.10997},
  volume={2},
  number={1},
  year={2023}
}

@article{li2024dawn,
  title={The dawn after the dark: An empirical study on factuality hallucination in large language models},
  author={Li, Junyi and Chen, Jie and Ren, Ruiyang and Cheng, Xiaoxue and Zhao, Wayne Xin and Nie, Jian-Yun and Wen, Ji-Rong},
  journal={arXiv preprint arXiv:2401.03205},
  year={2024}
}

@article{sun2023chatgpt,
  title={Is ChatGPT good at search? investigating large language models as re-ranking agents},
  author={Sun, Weiwei and Yan, Lingyong and Ma, Xinyu and Wang, Shuaiqiang and Ren, Pengjie and Chen, Zhumin and Yin, Dawei and Ren, Zhaochun},
  journal={arXiv preprint arXiv:2304.09542},
  year={2023}
}

@article{pradeep2023rankzephyr,
  title={RankZephyr: Effective and Robust Zero-Shot Listwise Reranking is a Breeze!},
  author={Pradeep, Ronak and Sharifymoghaddam, Sahel and Lin, Jimmy},
  journal={arXiv preprint arXiv:2312.02724},
  year={2023}
}

@inproceedings{hou2024large,
  title={Large language models are zero-shot rankers for recommender systems},
  author={Hou, Yupeng and Zhang, Junjie and Lin, Zihan and Lu, Hongyu and Xie, Ruobing and McAuley, Julian and Zhao, Wayne Xin},
  booktitle={European Conference on Information Retrieval},
  pages={364--381},
  year={2024},
  organization={Springer}
}

@article{wei2021finetuned,
  title={Finetuned language models are zero-shot learners},
  author={Wei, Jason and Bosma, Maarten and Zhao, Vincent Y and Guu, Kelvin and Yu, Adams Wei and Lester, Brian and Du, Nan and Dai, Andrew M and Le, Quoc V},
  journal={arXiv preprint arXiv:2109.01652},
  year={2021}
}

@article{crispino2023agent,
  title={Agent instructs large language models to be general zero-shot reasoners},
  author={Crispino, Nicholas and Montgomery, Kyle and Zeng, Fankun and Song, Dawn and Wang, Chenguang},
  journal={arXiv preprint arXiv:2310.03710},
  year={2023}
}

@article{liusie2023llm,
  title={LLM comparative assessment: Zero-shot NLG evaluation through pairwise comparisons using large language models},
  author={Liusie, Adian and Manakul, Potsawee and Gales, Mark JF},
  journal={arXiv preprint arXiv:2307.07889},
  year={2023}
}

@article{li2022self,
  title={Self-prompting large language models for zero-shot open-domain QA},
  author={Li, Junlong and Wang, Jinyuan and Zhang, Zhuosheng and Zhao, Hai},
  journal={arXiv preprint arXiv:2212.08635},
  year={2022}
}

@article{shi2023prompt,
  title={Prompt space optimizing few-shot reasoning success with large language models},
  author={Shi, Fobo and Qing, Peijun and Yang, Dong and Wang, Nan and Lei, Youbo and Lu, Haonan and Lin, Xiaodong and Li, Duantengchuan},
  journal={arXiv preprint arXiv:2306.03799},
  year={2023}
}

@article{machi2009literature,
  title={The literature review: Six steps to success},
  author={Machi, Lawrence A and McEvoy, Brenda T},
  year={2009},
  publisher={Corwin Press}
}

@article{arshed2015literature,
  title={The literature review},
  author={Arshed, Norin and Danson, Mike},
  journal={Research methods for business and management: a guide to writing your dissertation},
  pages={31--49},
  year={2015},
  publisher={Goodfellow Publishers}
}

@article{webster2002analyzing,
  title={Analyzing the past to prepare for the future: Writing a literature review},
  author={Webster, Jane and Watson, Richard T},
  journal={MIS quarterly},
  pages={xiii--xxiii},
  year={2002},
  publisher={JSTOR}
}

@article{yao2023tree,
  title={Tree of thoughts: Deliberate problem solving with large language models},
  author={Yao, Shunyu and Yu, Dian and Zhao, Jeffrey and Shafran, Izhak and Griffiths, Tom and Cao, Yuan and Narasimhan, Karthik},
  journal={Advances in neural information processing systems},
  volume={36},
  pages={11809--11822},
  year={2023}
}

@article{merton1968matthew,
  title={The Matthew effect in science: The reward and communication systems of science are considered},
  author={Merton, Robert K},
  journal={Science},
  volume={159},
  number={3810},
  pages={56--63},
  year={1968},
  publisher={American Association for the Advancement of Science}
}

@article{rossiter1993matilda,
  title={The Matthew Matilda effect in science},
  author={Rossiter, Margaret W},
  journal={Social studies of science},
  volume={23},
  number={2},
  pages={325--341},
  year={1993},
  publisher={SAGE Publications}
}

@article{kozlowski2024howard,
  title={The Howard-Harvard effect: Institutional reproduction of intersectional inequalities},
  author={Kozlowski, Diego and Monroe-White, Thema and Larivi{\`e}re, Vincent and Sugimoto, Cassidy R},
  journal={Journal of the Association for Information Science and Technology},
  volume={75},
  number={8},
  pages={869--882},
  year={2024},
  publisher={Wiley},
  doi={10.1002/asi.24931}
}

@article{lariviere2015oligarchy,
  title={The oligarchy of academic publishers in the digital era},
  author={Larivi{\`e}re, Vincent and Haustein, Stefanie and Mongeon, Philippe},
  journal={PloS one},
  volume={10},
  number={6},
  pages={e0127502},
  year={2015},
  publisher={Public Library of Science}
}

@book{sugimoto2018lariviere,
  title={Measuring research: What everyone needs to know},
  author={Sugimoto, Cassidy R and Larivi{\`e}re, Vincent},
  year={2018},
  publisher={Oxford University Press}
}


\end{document}